\documentclass[]{spie}  

\usepackage{amsmath,amsfonts,amssymb}
\usepackage{graphicx}
\usepackage[colorlinks=true, allcolors=blue]{hyperref}

\usepackage{caption}

\usepackage{tabularx}

\usepackage{mathrsfs} 

\usepackage{xcolor}

\usepackage{bbm}

\title{Observer study-based evaluation of a stochastic and physics-based method to generate oncological PET images}

\author[a]{Ziping Liu}
\author[b]{Richard Laforest}
\author[b]{Joyce Mhlanga}
\author[b]{Tyler J. Fraum}
\author[b]{Malak Itani}
\author[b]{Farrokh Dehdashti}
\author[b]{Barry A. Siegel}
\author[a,b]{Abhinav K. Jha}

\affil[a]{Department of Biomedical Engineering, Washington University in St. Louis, St. Louis, MO, USA}
\affil[b]{Mallinckrodt Institute of Radiology, Washington University School of Medicine, St. Louis, MO, USA}

\authorinfo{Corresponding author: Abhinav K. Jha (a.jha@wustl.edu)}

\begin{document} 

\begin{titlepage}
This manuscript has been accepted to SPIE Medical Imaging, February 15-19, 2021. Please use the following reference when citing the manuscript.

Liu, Z., Laforest, R., Mhlanga, J., Fraum, T. J., Itani, M., Dehdashti, F., Siegel, B. A., and Jha, A. K., ``Observer study-based evaluation of a stochastic and physics-based method to generate oncological PET images", Proc. SPIE Medical Imaging, 2021.

\end{titlepage}

\clearpage

\maketitle

\begin{abstract}

Objective evaluation of new and improved methods for PET imaging requires access to images with ground truth, as can be obtained through simulation studies. However, for these studies to be clinically relevant, it is important that the simulated images are clinically realistic. In this study, we develop a stochastic and physics-based method to generate realistic oncological two-dimensional ($2$-D) PET images, where the ground-truth tumor properties are known. The developed method extends upon a previously proposed approach {\cite{leung2020physics}}. The approach captures the observed variabilities in tumor properties from actual patient population. Further, we extend that approach to model intra-tumor heterogeneity using a lumpy object model. To quantitatively evaluate the clinical realism of the simulated images, we conducted a human-observer study. This was a two-alternative forced-choice (2AFC) study with trained readers (five PET physicians and one PET physicist). Our results showed that the readers had an average of $\sim 50\%$ accuracy in the 2AFC study. Further, the developed simulation method was able to generate wide varieties of clinically observed tumor types. These results provide evidence for the application of this method to $2$-D PET imaging applications, and motivate development of this method to generate $3$-D PET images.

\end{abstract}


\section{Introduction}

Positron emission tomography (PET) is a widely used imaging modality with multiple clinical applications, in particular for diagnosis and assessment of treatment response of cancers{\cite{zhu2011metabolic}}. Thus, several new and improved methods have been developed for oncological PET image reconstruction{\cite{tong2010image}}, segmentation{\cite{foster2014review}}, and quantification{\cite{tomasi2012importance}}. Objective evaluation and optimization of these methods typically requires knowledge of the corresponding ground truth. For instance, evaluation of segmentation methods typically requires knowledge of the ground-truth tumor boundaries. Similarly, optimization of PET imaging methods for clinical tasks using objective assessment of image quality (OAIQ) studies{\cite{barrett2013objective}} typically requires knowledge of the ground truth. Simulation studies provide a mechanism for such evaluation since the ground truth is known in these studies. Further, these studies provide the ability to model \textit{in vivo} anatomical and physiological properties of the patient, incorporate patient-population variability, model imaging-system physics, and generate multiple scan realizations of the same patient to evaluate repeatability. Even more importantly, this is all done \textit{in silico}, which is inexpensive and enables optimizing the method before conducting clinical studies. However, for these evaluation and optimization studies to be clinically relevant, the simulated PET images must be clinically realistic. 

In simulation-based studies to evaluate oncological PET methods, one set of studies uses synthetic phantoms, such as the NEMA phantom{\cite{nyflot2015quantitative}}. However, these phantoms have limited ability to model patient anatomy and physiology. Thus, to improve clinical realism, anthropomorphic phantom-based studies, such as using the XCAT phantom{\cite{segars20104d}}, have been conducted{\cite{lorsakul20144d}}. However, recent studies suggest that anthropomorphic phantoms may have limitations in modeling patient physiology{\cite{leung2020physics}}. Another limitation of existing simulation studies is that the tumor is typically modeled as a spherically shaped structure with limited incorporation of intra-tumor heterogeneity{\cite{nyflot2015quantitative,lorsakul20144d,cysouw2016accuracy}}. To incorporate variabilities in patient population and tumor models for more clinically relevant evaluation, Leung et al.{\cite{leung2020physics}} developed a simulation-based strategy that uses patient images as backgrounds and generates a wide variety of clinically observed tumor types. However, this strategy had limitations in modeling variabilities in intra-tumor heterogeneity. This heterogeneity is often observed in tumors, and methods to quantify this heterogeneity from PET images and evaluate its clinical predictive and prognostic value is a topic of intense research{\cite{cook2018challenges,mena2017}}. To model this intra-tumor heterogeneity more realistically, in this manuscript, we propose a lumpy model-based approach. Using this approach in conjunction with the simulation-based strategy, we develop a stochastic and physics-based method to generate clinically realistic PET images. 

The second contribution of this manuscript is in providing a theoretical premise for an observer-study-based framework to quantitatively evaluate the clinical realism of simulated images. An important goal of evaluating the clinical realism is that the simulation studies should accurately capture the variabilities in patient population. Further, ideally, the distribution of simulated images should match that of real images. Human-observer studies have been applied to evaluate the clinical realism of simulated images {\cite{chen2016validation,Ma2016}}. These studies account for the role of end users such as radiologists in clinical tasks. Human-observer studies are typically conducted using either rating-based methods or forced-choice-based methods. In this study, we consider specifically a forced-choice-based method, namely the two-alternative forced-choice (2AFC) study. We first provide a theoretical premise for this study to evaluate the realism of simulated images. We then apply the study to quantitatively evaluate the realism of the simulated PET images generated using our developed simulation method. 

\section{Methods}

In this section, we first describe the developed stochastic and physics-based method to generate simulated PET images. The theoretical background and methods for conducting the 2AFC study are subsequently provided. 

\subsection{Generating simulated PET images}
\label{sec:simul framework}

This study was conducted in the context of simulating the primary tumor in [$^{18}$F]fluorodeoxyglucose (FDG)-PET images of patients with lung cancer. The study was retrospective, used clinical imaging data, was approved by our institutional review board, and was HIPAA-compliant with a waiver of informed consent.

\begin{figure}[b!]
    \centering
    \includegraphics[width=\textwidth]{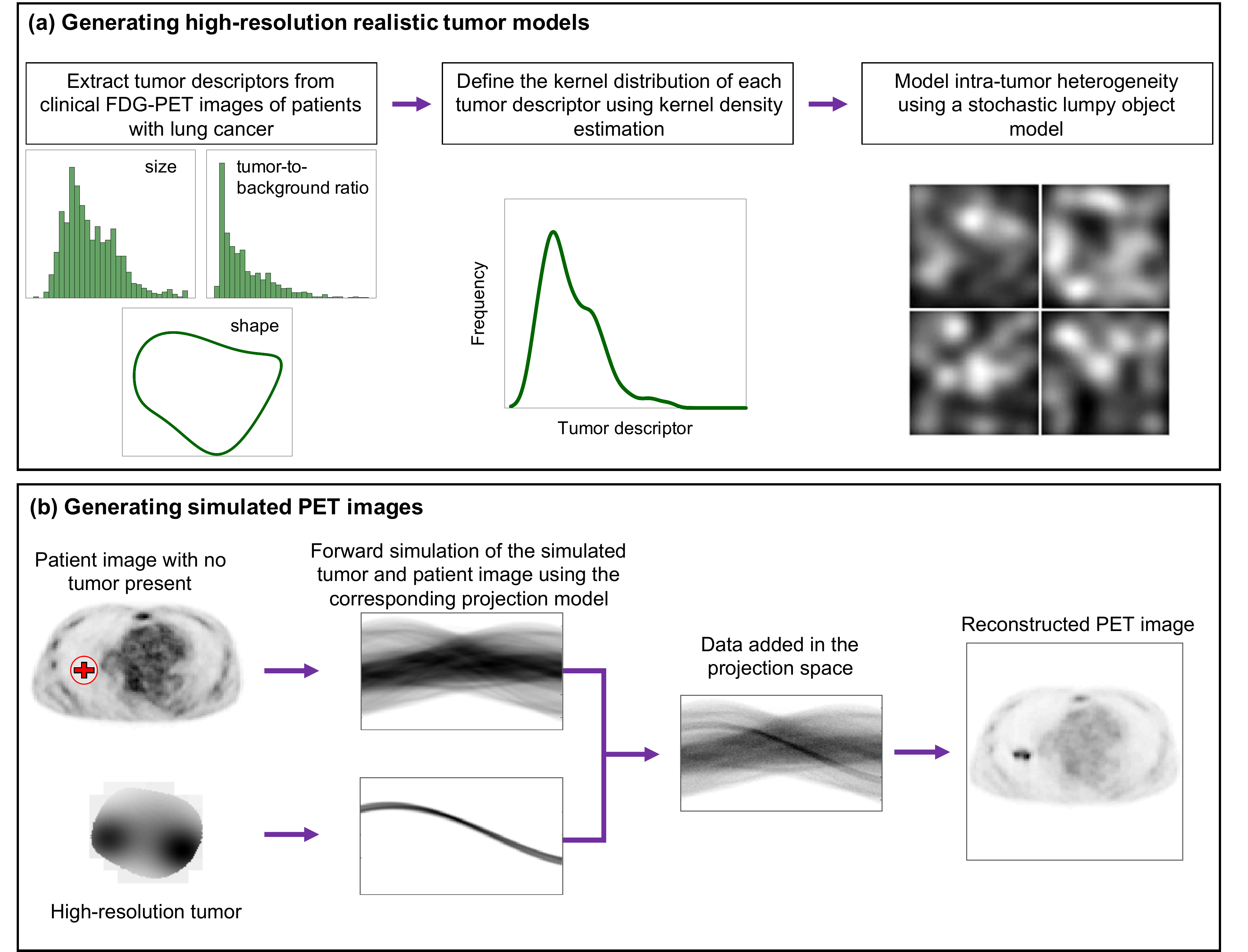}
    \captionsetup{justification=centering}
    \caption{Description of the developed method to first (a) generate high-resolution realistic tumor models with tumor properties extracted from clinical data and intra-tumor heterogeneity modeled using a stochastic lumpy object model, and then (b) generate simulated PET images using the patient images and tumor model.}
    \label{fig:generate_sim_img}
\end{figure}

The method is summarized in Fig.~\ref{fig:generate_sim_img}. In the first step, a realistic high-resolution tumor model was developed to capture the observed variabilities in tumor properties from an actual patient population (Fig.~\ref{fig:generate_sim_img}a). For this purpose, we advanced on a simulation-based strategy proposed by Leung et al.{\cite{leung2020physics}}. Briefly, tumor descriptors, including first- and second-order statistics for the shape, size, and tumor-to-background intensity ratio, were first extracted from clinical FDG-PET images of patients with lung cancer. Tumor shape was quantified by five harmonic elliptical Fourier shape descriptors{\cite{kuhl1982elliptic}}, and tumor size was quantified by diameter and volume. The distribution of each tumor descriptor was defined using kernel density estimation. The kernel distribution of each descriptor was then sampled, and from the sampled parameters, simulated tumors were generated. In Leung et al.{\cite{leung2020physics}}, necrosis within the tumor was modeled by assigning a lower intensity to the tumor core than the rim. We advanced on that approach to model intra-tumor heterogeneity more realistically. We used the observation that the tracer-uptake patterns within tumors can be modeled as a combination of lumps, where the lump locations, amplitudes, and sizes are random variables. More generally, it is suitable to characterize the tracer uptake within a tumor as a random process{\cite{henscheid2018physiological}}. Thus, the intra-tumor heterogeneity was modeled using a stochastic lumpy object model. This lumpy object model was inspired by the original lumpy background model{\cite{rolland1992effect}}, but with some adaptations to account for intra-tumor heterogeneity. Our lumpy object model was given by
\begin{equation}
    f(\mathbf{r}) = s(\mathbf{r})\sum_{n=1}^N \Lambda(\mathbf{r} - \mathbf{c}_n|a_n, \sigma_n) 
    = s(\mathbf{r})\sum_{n=1}^N \frac{a_n}{2 \pi \sigma_n^2} \mathrm{exp} \left( -\frac{|\mathbf{r} - \mathbf{c}_n|^2}{\sigma_n^2} \right),
\end{equation}
where $s(\mathbf{r})$ denotes the support for the tumor, $N$ denotes the total number of lumps, $\Lambda(\cdot)$ denotes the lump function, $\mathbf{r}$ denotes the spatial coordinate in two dimensions, and $\mathbf{c}_n$, $a_n$, and $\sigma_n$ denote the center, magnitude, and width of the $n^{\mathrm{th}}$ lump function, respectively. To model the tracer uptake as a random process, $\mathbf{c}_n$ was uniformly distributed within the support of tumor, and $a_n$ and $\sigma_n$ were uniformly distributed within a pre-defined range but appropriately scaled based on the clinically extracted values of the tumor-to-background intensity ratios.

Through this strategy, high-resolution simulated tumors with known ground-truth properties were generated. Note that the ground truth was not needed for the background. To ensure the clinical realism of tumor background and model inter-patient variability, existing patient images containing lung cavities but with no tumor present were selected as tumor background. To ensure that the simulated tumors only appear at visually realistic locations within the lung cavities, tumor locations were manually identified in advance. The simulated tumors were then randomly generated and placed at these locations.

In the second step (Fig. {\ref{fig:generate_sim_img}}b), forward projections for the simulated tumor and patient background were generated using a PET simulation software {\cite{leung2020physics}}. The high-resolution simulated tumor and low-resolution patient background were passed through corresponding projection models to obtain the projection data. Similar to Ma et al.{\cite{Ma2016}}, adding the data in the projection space and then performing reconstruction helped incorporate the impact of noise texture on the tumor appearance in the reconstructed image. The reconstruction was performed using a $2$-D ordered subset expectation maximization (OSEM) algorithm. Detailed simulation and reconstruction parameters are provided in Table \ref{tab:simul_paramters}.

\begin{table}[h]
    \centering
    \normalsize
    \captionsetup{justification=centering}
    \caption{Technical acquisition and reconstruction parameters of the PET system. (FWHM: full-width-half-maximum)}
    \begin{tabularx}{0.9\textwidth}{
    | >{\hsize=1\hsize\centering\arraybackslash}X
    | >{\hsize=1\hsize\centering\arraybackslash}X |}
    \hline
    Parameters & Values \\
    \hline
    Transaxial field of view & 684 $\mathrm{mm}$ \\
    \hline
    Pixel size & 4.07 $\mathrm{mm}$ $\times$ 4.07 $\mathrm{mm}$ \\
    \hline
    Reconstruction method & OSEM\\ 
    \hline
    Subsets & 21\\ 
    \hline
    Iterations & 2 \\ 
    \hline
    FWHM @ 1 $\mathrm{cm}$ & 5 $\mathrm{mm}$ \\
    \hline
    \end{tabularx}
    \label{tab:simul_paramters}
\end{table}

\subsection{Evaluating the realism of the simulated PET images} 

To quantitatively evaluate the clinical realism of the simulated PET images, we conducted a 2AFC study. We first provide the theoretical background for conducting this study. It is well known that computing the probability of an observer making a correct assignment in the 2AFC study is the same as computing the AUC of that observer. We first consider a hypothetical scenario, where an ideal observer can be constructed to discriminate between the real and simulated images. Following the treatment in Barrett et al. {\cite{barrett1998objective,barrett2013foundations}}, but in the context of the task of discriminating between real and simulated images, we show that an ideal-observer AUC of $0.5$ on this task leads to the inference that the distribution of simulated images exactly matches that of real images. However, an ideal observer is challenging to implement for this discrimination task. Thus, we instead use a practical but rigorous option, namely evaluation by trained human observers with multiple years of experience in reading PET scans. We first present the theoretical background for conducting this observer study.

\subsubsection{Theoretical background}
\label{sec:2AFC justification}

Denote the sets of simulated and real PET images by $\mathbf{\hat{f}}$ and $\mathbf{\hat{f}'}$, each in M-dimensional space. Consider two hypotheses ${H_1}$ and ${H_2}$, where ${H_1}$ and ${H_2}$ refer to the class of simulated and real PET images, respectively. Denote the conditional probability distribution of the observed data $\mathbf{\hat{f}}$ under hypothesis $j$ by $\mathrm{pr(\mathbf{\hat{f}}|\textit{H}_j)}$. For convenience in notation, we define $q_j(\mathbf{\hat{f}}) \equiv \mathrm{pr(\mathbf{\hat{f}}|\textit{H}_j)}$.  
In the 2AFC study, an observer is presented with two sets of images $\mathbf{\hat{f}}$ and $\mathbf{\hat{f}'}$ such that $\mathbf{\hat{f}}$ is sampled from $q_1(\mathbf{\hat{f}})$ and $\mathbf{\hat{f}'}$ is sampled from $q_2(\mathbf{\hat{f}}')$. The observer is then asked to select the image that they think is the real PET image.

The observer computes two test statistics $\theta(\mathbf{\hat{f}})$ and $\theta(\mathbf{\hat{f}'})$ and assigns the image that has the higher test statistic to ${H_2}$. The assignment is correct if $\theta(\mathbf{\hat{f}'}) > \theta(\mathbf{\hat{f}})$. Thus, the probability of a correct assignment can be computed as follows:
\begin{equation}
    \mathrm{pr}\Big[\theta(\mathbf{\hat{f}'}) > \theta(\mathbf{\hat{f}})\Big] 
    = \int_\infty d^M \mathbf{\hat{f}}  \ q_1(\mathbf{\hat{f}}) \ \int_\infty d^M \mathbf{\hat{f}'} \ q_2(\mathbf{\hat{f}'}) \ \mathrm{step}\Big[\theta(\mathbf{\hat{f}'}) - \theta(\mathbf{\hat{f}})\Big],
    \label{eq: prob of correct decision}
\end{equation}
where $\mathrm{step}(\cdot)$ denotes the step function. As shown in Barrett and Myers{\cite{barrett2013foundations}}, Eq. \eqref{eq: prob of correct decision} is the same as the equation to compute AUC for an arbitrary test statistic with unknown probability law. Thus, the percentage of times that an observer correctly identifies the real PET image is equivalent to the AUC of that observer. 

We now consider the special case of an ideal observer. An ideal observer can be defined as a decision strategy that computes the likelihood ratio $ \frac{q_2(\mathbf{\hat{f}})}{q_1(\mathbf{\hat{f}})}$ and compares it to a threshold. This likelihood ratio is considered as the optimal discriminant function and is a sufficient statistic that contains all the information needed to perform the discrimination task. 
For this ideal observer, we can use the concept of the likelihood-generating function{\cite{barrett1998objective}} to derive the relationship between the AUC and $q_1(\mathbf{\hat{f}})$ and $q_2(\mathbf{\hat{f}})$.  Essentially, this likelihood-generating function, defined as $G(\cdot)$, provides a lower bound for the AUC as follows:
\begin{equation}
    \mathrm{AUC} \geq 1 - \frac{1}{2} \mathrm{exp} \left[ - \frac{1}{2} G(0) \right],
\label{eq: ideal-observer AUC lower bound}
\end{equation}
where $G(0)$ is the likelihood-generating function evaluated at origin. $G(0)$ can further be expressed in terms of $q_1(\mathbf{\hat{f}})$ and $q_2(\mathbf{\hat{f}})$, i.e.
\begin{equation}
G(0) = -4 \ \mathrm{log} \left[ \int_\infty d^M \mathrm{\mathbf{\hat{f}}} \sqrt{q_1(\mathbf{\hat{f}})q_2(\mathbf{\hat{f}})} \right].
\label{eq: G(0)}
\end{equation}

From Eq.~\eqref{eq: ideal-observer AUC lower bound}, we see that a lower bound of $\mathrm{AUC}=0.5$ is achieved by setting $G(0) = 0$. From Eq.~\eqref{eq: G(0)}, $G(0) = 0$ is achieved when $q_1(\mathbf{\hat{f}}) = q_2(\mathbf{\hat{f}})$. Thus, an ideal-observer AUC of $0.5$ leads to the inference that the distributions of real and simulated images exactly match. 

Implementation of the ideal observer would require knowledge of the probability distribution of likelihood ratio under both hypotheses. While this is not feasible in our study, we use a practical but rigorous alternative, namely a set of trained human observers with multiple years of experience in reading PET scans. Specifically, five nuclear medicine radiologists and one PET physicist participated in the observer study. Our goal is to have an observer that that sets an upper limit to the performance of any available human observer. If the trained human observer obtains an AUC of $0.5$, we may infer that the distribution of the simulated images is close to that of the real images. Thus, our 2AFC study provides a rigorous and practical mechanism to validate the clinical realism of the simulated images. We next describe the design of this observer study to evaluate the clinical realism of the PET images generated using our simulation method. 

\subsubsection{Experimental design} \label{sec:2AFC protocol}
\begin{figure}[h!]
    \centering
    \includegraphics[width=0.9\textwidth]{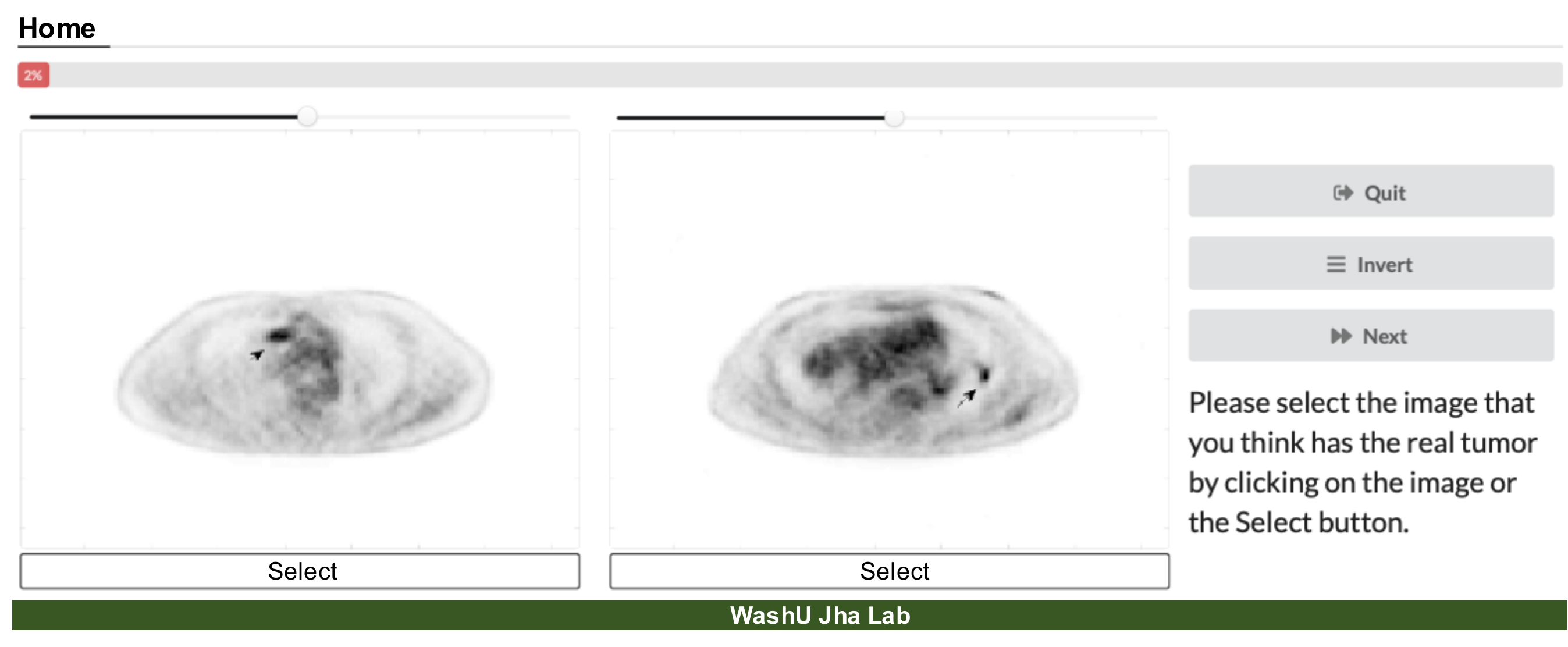}
    \caption{Interface of the web-based application presented to readers in the 2AFC study.}
    \label{fig:2-AFC}
    \captionsetup{justification=centering}
\end{figure}

To conduct the 2AFC study, we developed a web-based app (Fig.~\ref{fig:2-AFC}). During the study, the trained readers were shown two images side-by-side at a time, one a real patient image sampled from $q_2(\mathbf{\hat{f}})$ and the other a simulated PET image sampled from $q_1(\mathbf{\hat{f}})$ using our simulation method. As described in Sec. \ref{sec:2AFC justification}, the readers were instructed that the task was to identify the real PET image that they thought had the real tumor. The tumor location was shown in the images to ensure that the readers were focusing on the tumor-realism task, and not implicitly treating this as a tumor-detection task. To facilitate a robust observer study, the app incorporated functionalities provided by clinical software, including the option to invert the image intensities and adjust the image contrast. Further, the app used MySQL to manage the readers' records, easing data collection and analysis. 

Six trained readers, which included five board-certified radiologists with specialization in nuclear medicine and many years of experience in PET (B.A.S., F.D., J.C.M., T.J.F., M.I.) and one experienced nuclear-medicine physicist (R.L.), participated as readers in this study. The readers performed this test for $50$ image pairs. We computed the fraction of times that each reader correctly identified the patient image. As shown in Sec. \ref{sec:2AFC justification}, a percent accuracy close to 50\% for a well-trained reader suggests a high similarity between the distributions of the real and simulated images.

\section{RESULTS}
\label{sec:results}

\begin{figure}[h]
    \centering
    \includegraphics[width=0.9\textwidth]{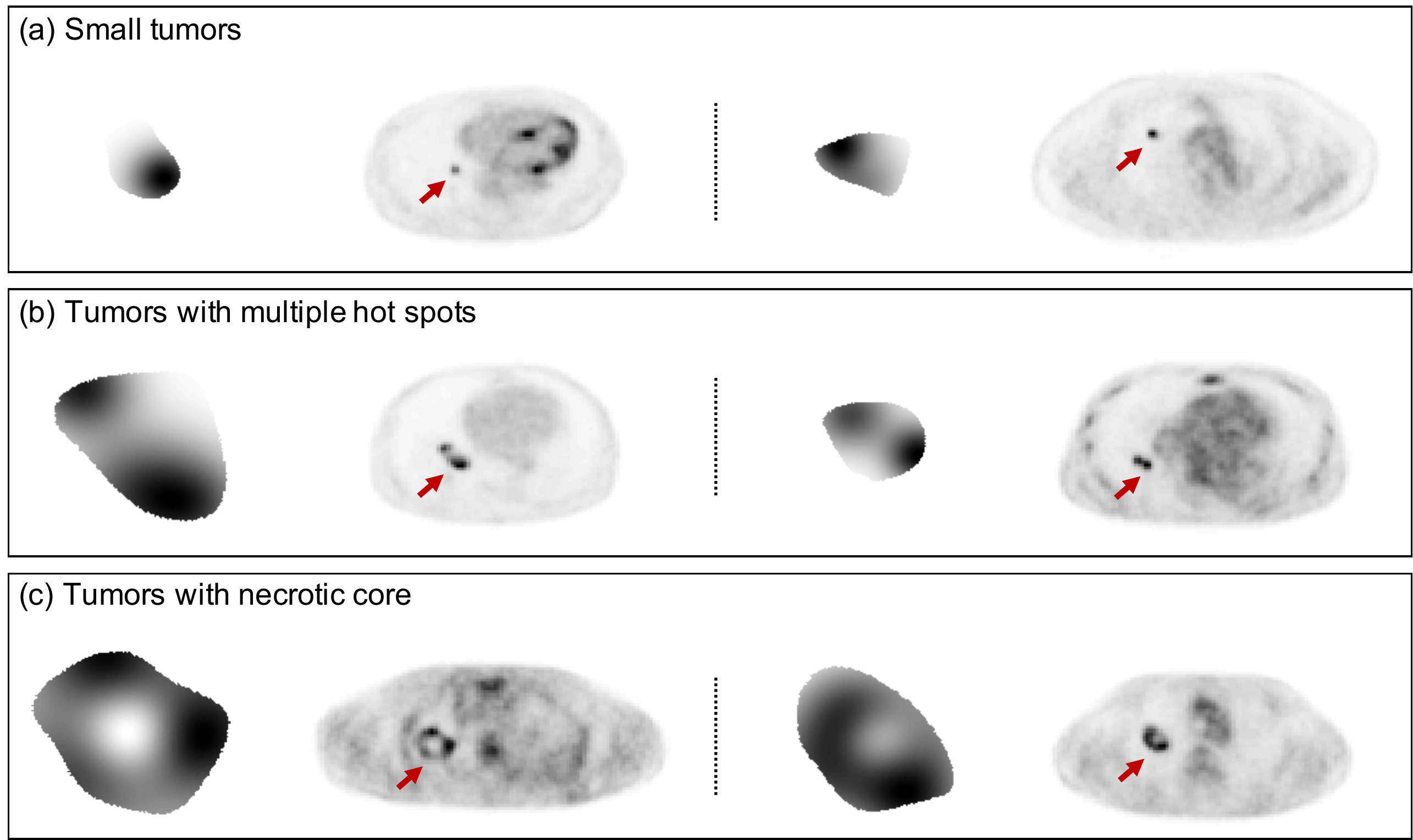}
    \caption{Representative high-resolution tumor models with different tumor types and the corresponding reconstructed PET images. Tumor locations in the reconstructed images are marked by the arrows.}
    \label{fig:representative_tumors}
    \captionsetup{justification=centering}
\end{figure}

Fig.~\ref{fig:representative_tumors} shows the representative simulated images generated using our simulation method (Sec.~\ref{sec:simul framework}). These images demonstrate that the method can generate a wide variety of clinically observed tumor types, including (a) small tumors, (b) tumors with multiple hot spots, and (c) tumors with necrotic cores.

\begin{figure}[h!]
    \centering
    \includegraphics[width=0.9\textwidth]{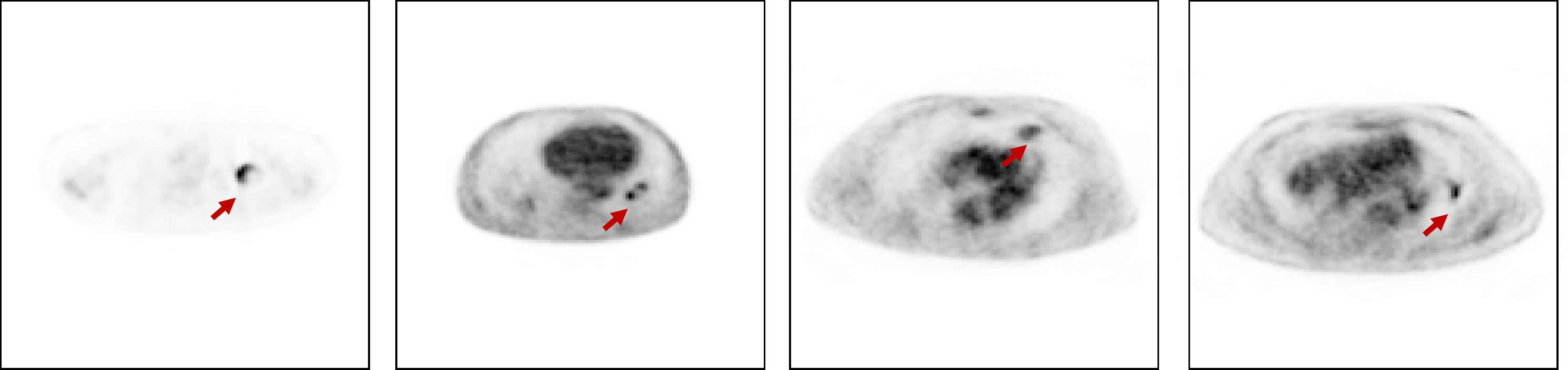}
    \caption{Representative simulated images incorrectly identified by at least 3 readers. Tumor locations are marked by the red arrows.}
    \label{fig:misidentification}
    \captionsetup{justification=centering}
\end{figure}

Table \ref{tab:2AFC_result} lists the results of the 2AFC study. Each trained reader identified the real images accurately in only approximately $50\%$ of the cases, suggesting that the distribution of the simulated images closely matches that of the real images (Sec. \ref{sec:2AFC justification}). Examples of simulated images that were incorrectly identified by at least half of the readers are shown in Fig.~\ref{fig:misidentification}. Overall, these results demonstrate that the simulated images generated using the developed simulation method are highly realistic.

\begin{table}[h]
    \centering
    \normalsize
    \captionsetup{justification=centering}
    \caption{Percent accuracy for each trained reader participating in the 2AFC test. (NM: nuclear medicine)}
    \begin{tabularx}{0.9\textwidth}{
    | >{\hsize=1\hsize\centering\arraybackslash}X
    | >{\hsize=1\hsize\centering\arraybackslash}X|}
    \hline
    &  Percent accuracy \\
    \hline
    NM physician 1 & 44\%  \\ 
    \hline
    NM physician 2 & 50\% \\
    \hline
    NM physician 3 & 58\% \\ 
    \hline
    NM physician 4 & 58\% \\ 
    \hline
    NM physician 5 & 44\% \\ 
    \hline
    NM physicist & 58\% \\ 
    \hline
    \end{tabularx}
    \label{tab:2AFC_result}
\end{table}

\section{DISCUSSIONS}
\label{sec:discussions}

In this manuscript, we first developed a stochastic and physics-based method to generate 2-D oncological PET images. Generation of realistic images is valuable for evaluating imaging methods for clinical tasks. For this purpose, several techniques have been developed to generate synthetic medical images. These include traditional data augmentation techniques, such as translation, scaling, shear, and rotation. However, these techniques fundamentally produce highly correlated images\cite{shin2018medical} and do not show an improvement in training performance in certain tasks{\cite{leung2020physics}}. Another approach to image generation is the use of generative adversarial network (GAN){\cite{goodfellow2014generative}}. GAN-based image generation has shown promise in multiple imaging modalities{\cite{shin2018medical,yi2019generative}}. However, such technique suffers from limitations of training instability{\cite{yi2019generative}} and requirement of large-scale training data. In addition, in the context of lung cancer, definition of ground-truth tumor properties is not well defined unlike our simulation method. Further, GAN-based techniques do not directly exploit the imaging physics and do not incorporate the variability in instrumentation{\cite{leung2020physics}}. In contrast, our method does not suffer from these limitations.

High realism of our simulated PET images, as quantitatively validated using the 2AFC study, motivates the application of our method to a broader range of quantitative evaluation studies. These include evaluation of imaging methods for segmentation tasks. For example, Liu et al.{\cite{liu2020estimation,liu2021estimationbased}} developed a deep-learning-based estimation approach to PET segmentation that estimates the tumor-fraction area within each pixel in a 2-D PET image. Our simulation method provides a mechanism to objectively evaluate the performance of this segmentation approach using clinically realistic simulation studies, where the ground-truth tumor boundaries are known. Our method can also be used to evaluate other imaging methods for metric quantification. In this context, simulation studies have been used to evaluate the performance of PVE correction techniques in PET. Existing simulation-based evaluation studies {\cite{bettinardi2014pet}} typically assume simplistic tumor models{\cite{cysouw2017impact}} such as spherically shaped tumors, and thus do not incorporate variability in actual patient population. The ability of our method to generate wide varieties of clinically observed tumor types  will make the evaluation studies more clinically relevant. Additionally, the method can be used to evaluate imaging methods for OAIQ-based studies in detection{\cite{gifford2014efficient,yu2020ai}} and quantification{\cite{jha2016no,jha2017practical,liu2020no,jha2017no}} tasks. Further, our proposed lumpy model-based approach to model intra-tumor heterogeneity could be used to evaluate image-reconstruction methods, as well as methods to quantify intra-tumor heterogeneity from PET images {\cite{nyflot2015quantitative,mena2017,mena2017value}}. In all these studies, access to clinically realistic tumor models will make the studies even more clinically relevant.

To conduct the 2AFC study, we developed a web-based app. The goal was to provide a mechanism to increase the flexibility of conducting this study. The web-based app eliminates the need to have the readers participate in the study on site.  In addition, this app provides functionalities to create a more familiar user interface design as observed in common clinical software. All these features increase the rigor and clinical relevance of the 2AFC study. This web app design can be naturally extended to conduct the 2AFC study for other image-simulation methods such as GAN-based approaches and is generalizable for other imaging modalities. Pending necessary permissions, we will publish this app on GitHub for wider usage by the image-science community.

Limitations of our study include the fact that the developed simulation method generates 2-D tumor models on transaxial image slices. While the method is less computationally expensive in 2-D tumor modeling, developing 3-D tumor models is important for incorporating the whole tumor features and thus is an important research area. Another area of future work is to simulate the PET physics and system instrumentation even more accurately, using approaches such as GATE {\cite{jan2004gate}}. One limitation in the observer study design is that the readers are typically trained on the task of detecting the tumor and not of discriminating the simulated tumor from the real tumor. One strategy to address this issue is to present examples of simulated images and real images to the readers prior to the 2AFC study and thus train them on this discrimination task. Another limitation of our study is that the choice of parameters for the lumpy object model to generate intra-tumor heterogeneity was based on visual inspection and not quantitatively obtained. In this context, several methods have been developed to fit statistical models of object based on image data {\cite{kupinski2003experimental,samei2019design}}. Thus, extending our method to statistically fit lumpy object model using patient data provides a mechanism to address this limitation.

\section{CONCLUSION}
\label{sec:conclusion}

In this manuscript, we quantitatively evaluated a stochastic and physics-based method to generate 2-D oncological PET images with known ground-truth tumor properties. A trained-reader-based observer study demonstrated that the method yielded highly realistic simulated images. In addition, the method demonstrates the ability to generate a wide variety of clinically observed tumor types, including tumors with complex intra-tumor heterogeneity. These results motivate the application of the method to a broader range of clinically relevant quantitative evaluation studies. Further, the theoretical premise for the observer study provides a foundation for the use of such observer-based studies to evaluate the clinical realism of images generated using other simulation-based approaches.

\section*{ACKNOWLEDGEMENTS}
Financial support for this work was provided by the Department of Biomedical Engineering and the Mallinckrodt Institute of Radiology at Washington University in St. Louis and an NVIDIA GPU grant. We also thank Qiye Tan for the help with developing the web-based app for the observer study. 

\bibliography{my_bib} 
\bibliographystyle{spiebib} 

\end{document}